\documentclass[a4paper]{article}
\usepackage{epsfig,amsmath,amsfonts,amssymb,setspace,multirow,textcomp}
\usepackage[T1]{fontenc}
\makeatletter
\renewcommand{\@oddhead}{\textit{Advances in Astronomy and Space Physics} \hfil}
\renewcommand{\@evenfoot}{\hfil \thepage \hfil}
\renewcommand{\@oddfoot}{\hfil \thepage \hfil}
\makeatother

 \renewenvironment{thebibliography}[1]{\begin{oldthebibliography}{#1}
 \setlength{\parskip}{0ex}\setlength{\itemsep}{0ex}}{\end{oldthebibliography}}

\usepackage{color,xspace,url}
\newcommand{\xmm}{\textit{XMM-Newton}\xspace}
\newcommand{\eq}[1]{\begin{equation} #1 \end{equation}}

\begin{document}
\fontsize{11}{11}\selectfont 
\title{Identification of the $\sim$3.55~keV emission line candidate objects across the sky}
\author{\textsl{D.\,O.~Savchenko$^{1}$, D.\,A.~Iakubovskyi$^{1}$}}
\date{\vspace*{-6ex}}
\maketitle
\begin{center} {\small $^{1}$Bogolyubov Institute of Theoretical Physics,
Metrologichna str. 14-b, 03680, Kyiv, Ukraine\\
{\tt dsavchenko@bitp.kiev.ua}}
\end{center}

\begin{abstract}
Emission line at the energy $\sim$3.55~keV detected in different galaxies and galaxy 
clusters has caused a lot of discussion in high-energy astrophysics and particle physics communities.
To reveal the origin of the line, we analyzed publicly available observations of MOS cameras from
\xmm\ cosmic observatory --
the instrument with the largest sensitivity for narrow faint X-ray lines -- previously combined 
in X-ray sky maps. Because of extremely large timescale needed for detailed analysis, 
we used the wavelet method instead. Extensive simulations
of the central part of Andromeda galaxy are used to check the validity of this method. The resulting list 
of wavelet detections now contain 235 sky regions. This list will be used in future works 
for more detailed spectral analysis.\\[1ex]
\end{abstract}

\section*{\sc Introduction}

\indent \indent The new narrow emission line at $\sim$3.55~keV reported in February 2014 from different 
stacks of galaxy clusters~\cite{Bulbul:14a}, the Andromeda galaxy and the Perseus 
galaxy cluster~\cite{Boyarsky:14a}, still remains unexplained, 
see reviews~\cite{Iakubovskyi:14,Iakubovskyi:15c} for details. 
Although the standard astrophysical explanation due to enhanced K~XVIII emission lines at $\sim$3.51~keV
proposed in~\cite{Jeltema:14a} remains possible~\cite{Iakubovskyi:15a}, 
subsequent measurements of the new line strength in the Galactic 
Centre~\cite{Boyarsky:14b,Lovell:14} and in the nearby dark matter-dominated galaxy 
clusters~\cite{Iakubovskyi:15b} argue for physics beyond the Standard Model, 
presumably in a form of radiatively decaying dark matter.
Further studies with future high-resolution imaging spectrometers, such as Soft X-ray Spectrometer 
(SXS) on-board forthcoming \textit{Astro-H} mission~\cite{Mitsuda:14} 
and \textit{Micro-X} sounding rocket experiment~\cite{Figueroa-Feliciano:15} will be able to determine 
the line origin~\cite{Bulbul:14a,Speckhard:15,Iakubovskyi:15a} in the nearest future.

Selection of the best follow-up targets is in progress. According to~\cite{Boyarsky:06f}, 
European Cosmic Imaging Camera (EPIC)~\cite{Turner:00,Strueder:01} 
on-board \xmm\ cosmic mission~\cite{Jansen:01} is the most sensitive existing instrument
in order to search the narrow faint X-ray line.
Recent measurement reported in~\cite{Iakubovskyi:15b} based on archival \xmm/EPIC data on galaxy clusters 
with the largest expected decaying dark matter signal doubles the number of the 
new line detections compared with Table~1 in~\cite{Iakubovskyi:14}. The result reported in~\cite{Iakubovskyi:15b}
encourages to further search for the $\sim$3.55~keV line. 
An example of the dataset to explore is the \xmm/EPIC sky map of~\cite{Savchenko:14} 
what contains about 4000 individual observations (80~Ms of cleaned exposure) by the EPIC/MOS cameras of \xmm.

Because detailed analysis of thousands of individual objects becomes extremely challenging task,
in this paper we propose the selection procedure of potential $\sim$3.55~keV line targets 
based on wavelet analysis.

\section*{\sc Methods}
\label{sec:methods}

Usually, wavelets in astrophysics are used for point sources detection~\cite{Slezak:94,Damiani:97,Freeman:01} 
and periodicity analysis~\cite{Foster:96,Krivova:02,Ellis:02}, but they can also be used for
search of local spectral inhomogeneities. 
As a simplified idea, we consider continuous spectrum with narrow line centered on bin $E_i$ with flux $F_i$
in $i$-th energy bin. By calculating flux residuals with respect to adjacent bins, 
\eq{\Delta F_i = {F_i}-\frac12\left(F_{i-1}+F_{i+1}\right),} 
we note that continuum contribution will roughly cancel while the line component localized in $i$-th bin
will not. Sliding along the spectral range of our interest and calculating the largest $\Delta F_i$, 
one can determine the line position $E_0$. Another important 
quantity -- line significance $S$ -- can be estimated as
\eq{S(E_0) = \frac{{F_i}-\frac12\left(F_{i-1}+F_{i+1}\right)}{\sqrt{F_i + \frac12\left(F_{i-1}+F_{i+1}\right)}}.
\label{eq:step-function}}
Its generalization on arbitrary wavelet function $\psi(t)$ is straightforward:
\eq{S(E_0) = \frac{ \int dE\psi(E-E_0)F(E)}{\sqrt{\int dE|\psi(E-E_0)|F(E)}}.\label{eq:significance-estimate}}

\begin{figure}[!h]
\centering
\begin{minipage}[t]{.45\linewidth}
\centering
\epsfig{file = 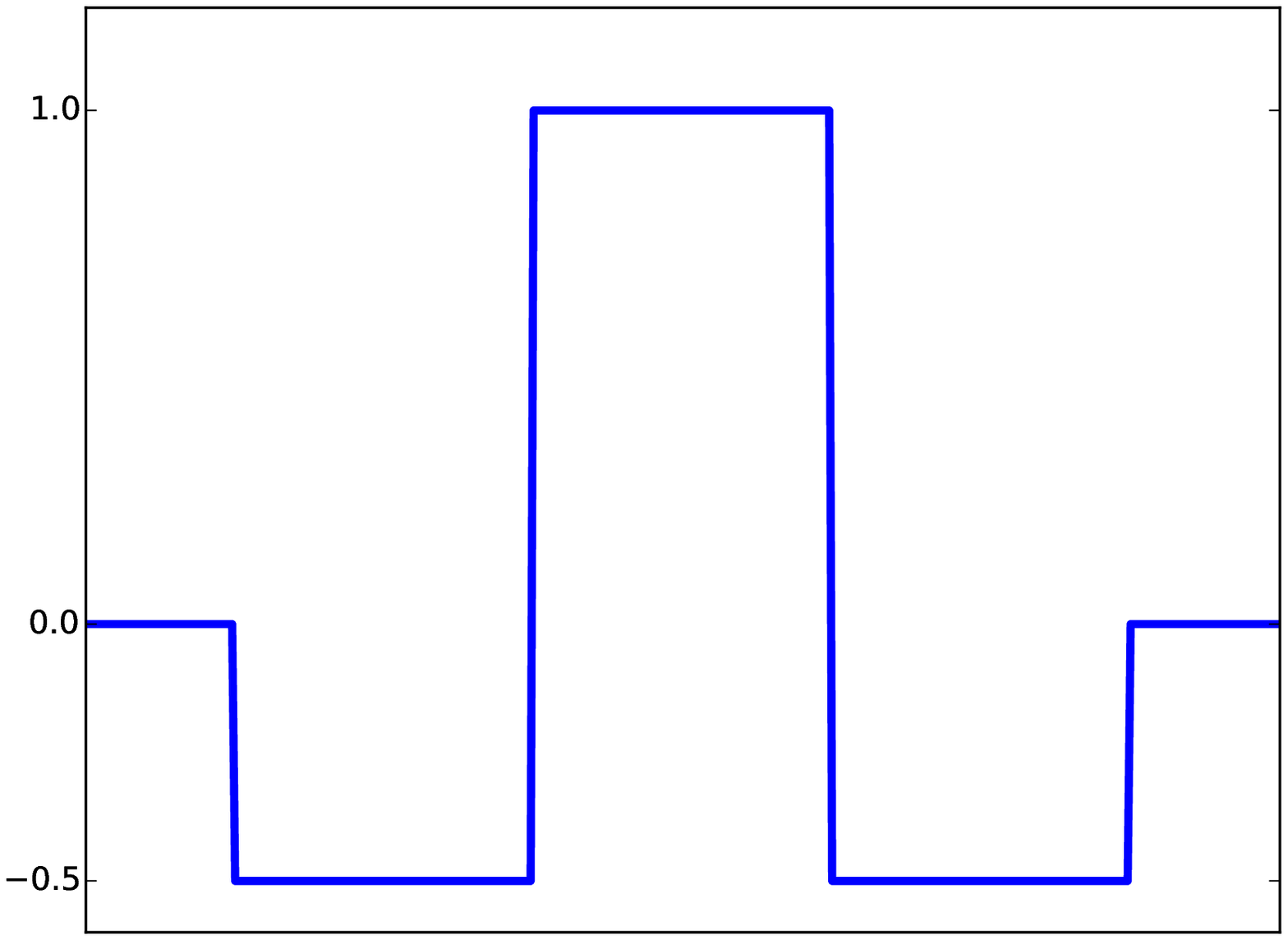,width = .85\linewidth}
\caption{Step function wavelet.}\label{fig:stepwl}
\end{minipage}
\hfill
\begin{minipage}[t]{.45\linewidth}
\centering
\epsfig{file = 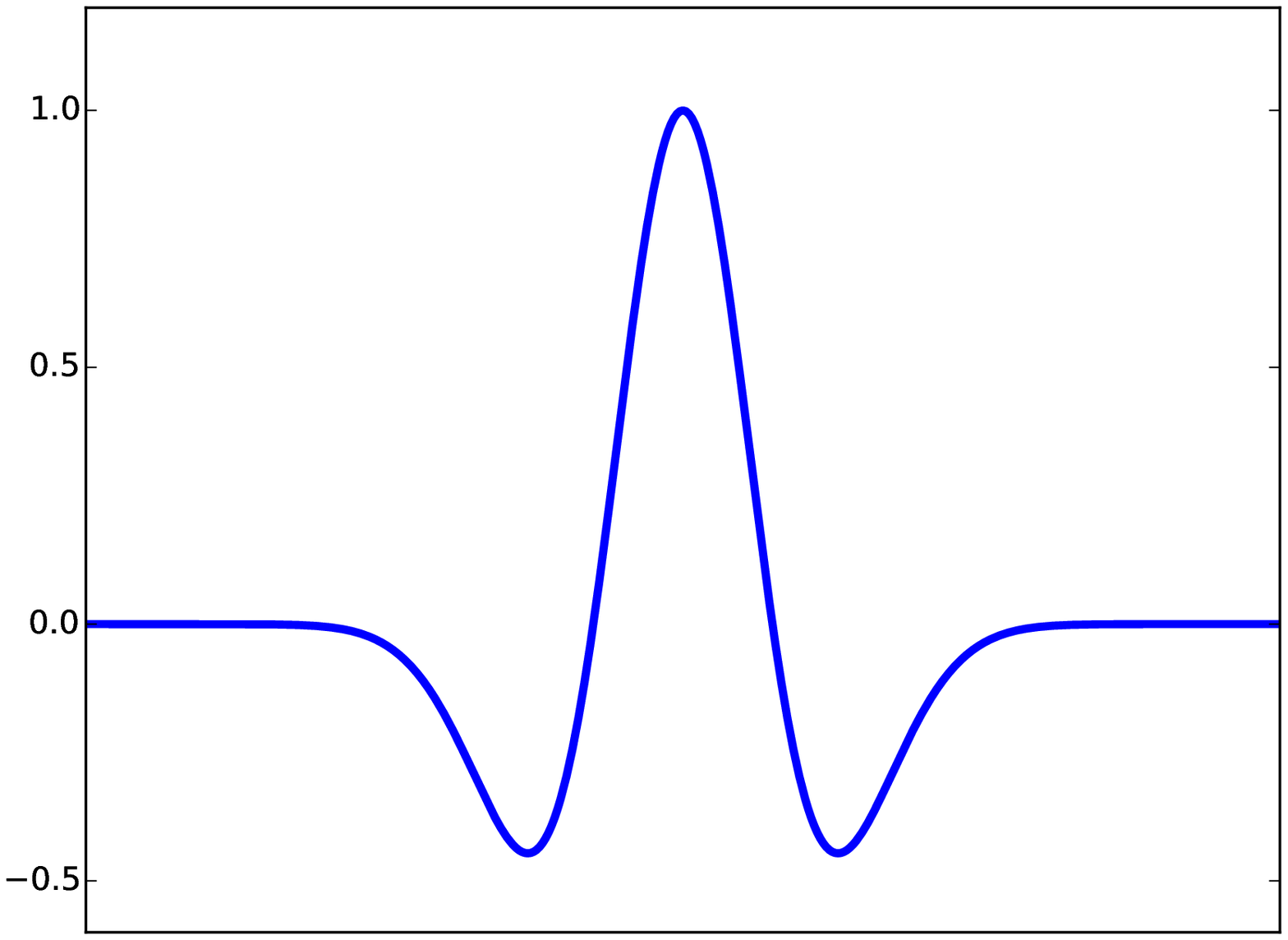,width = .85\linewidth}
\caption{``Mexican hat'' wavelet.}\label{fig:mhwl}
\end{minipage}
\end{figure}

For our analysis, we used 2 types of wavelet functions $\psi$:
\begin{itemize}
 \item step function wavelet \eq{\psi(t) = \frac32\,\theta\left(t+\frac{W}{2}\right) - \frac32\,\theta\left(t-\frac{W}{2}\right) - 
 \frac12\,\theta\left(t+\frac{3W}{2}\right)+\frac12\,\theta\left(t-\frac{3W}{2}\right)\label{eq:step}}
(where $\theta(t)$ is Heaviside function, $W$ is the width of the wavelet) 
used in Eq.~\ref{eq:step-function}, see Fig.~\ref{fig:stepwl};
 \item ``mexican hat'' wavelet 
 \eq{\psi(t)=\left( 1-\frac{t^2}{\sigma^2}\right)\exp\left(-\frac{t^2}{2\sigma^2}\right),\label{eq:mexican}} 
 see Fig.~\ref{fig:mhwl}.
\end{itemize}

\begin{figure}[!h]
\centering
\begin{minipage}[t]{.45\linewidth}
\centering
\epsfig{file = 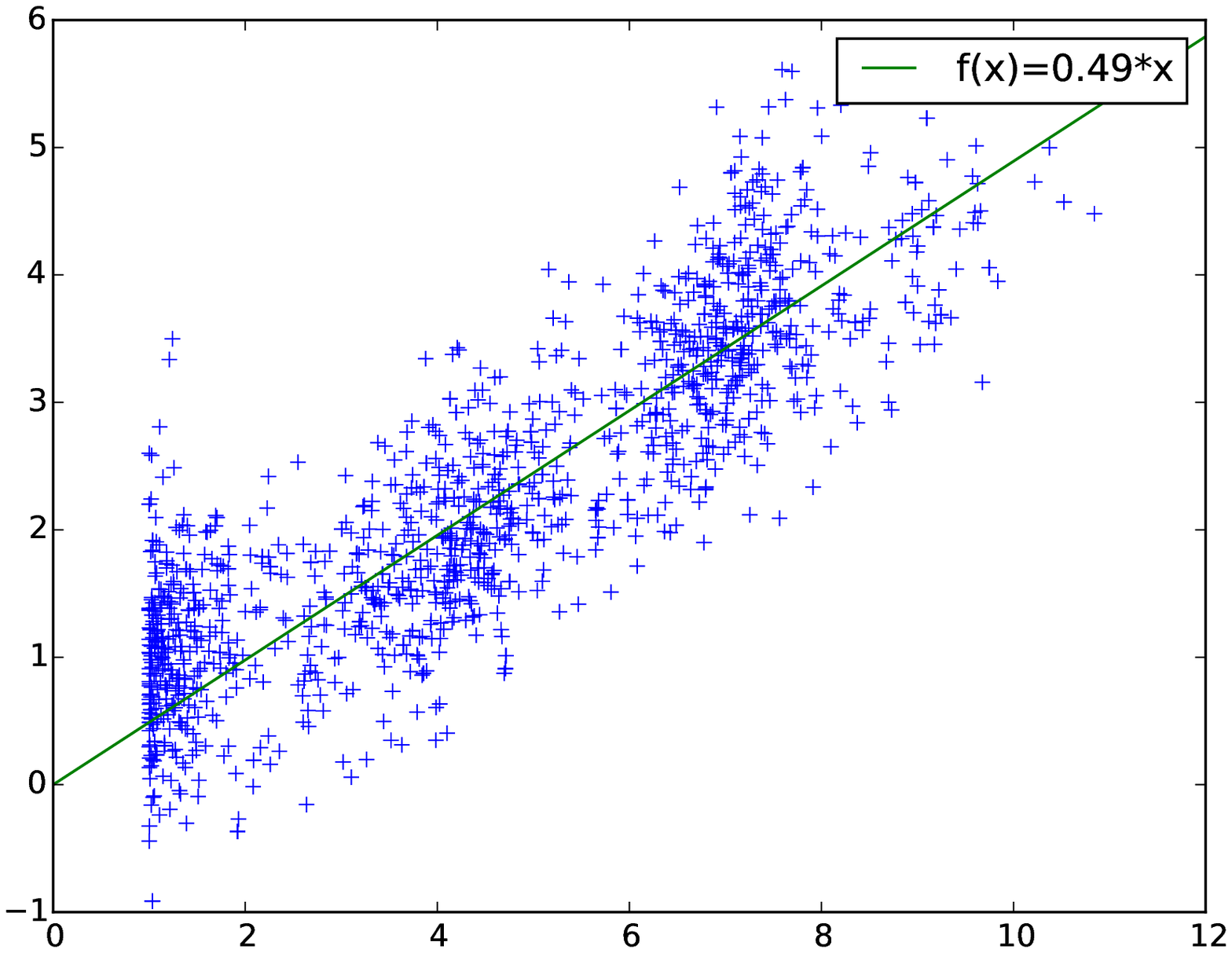,width = .85\linewidth}
\caption{Dependence of our step function wavelet significance estimator $S$ (on Y-axis) 
from the local line significance $\Sigma$ (on X-axis), see text. Wavelet width is set to $W = 120$~eV.}
\label{fig:stwsign120}
\end{minipage}
\hfill
\begin{minipage}[t]{.45\linewidth}
\centering
\epsfig{file = 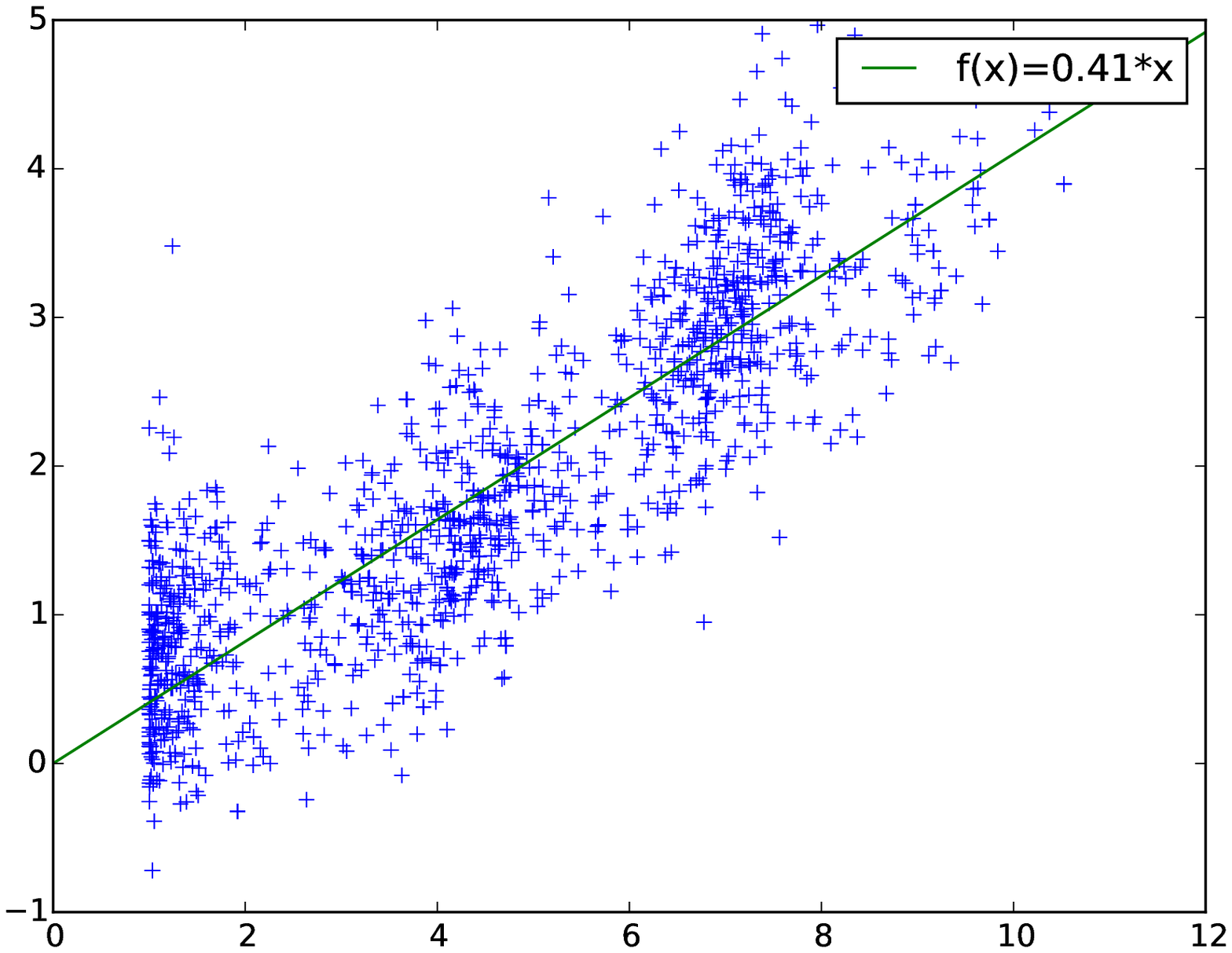,width = .85\linewidth}
\caption{The same as in Fig.~\protect\ref{fig:stwsign120} but for ``mexican hat'' wavelet 
with $\sigma = 60$~eV.}\label{fig:mhwsign60}
\end{minipage}
\end{figure}

In practice, using wavelet method should decrease the significance of the detected line.
The reasons are the presence of non-negligible instrumental emission lines (such as Potassium K$\alpha$ line at 3.31~keV
and Calcium K$\alpha$ line at 3.69~keV), complexes of astrophysical lines emitted by hot plasma
(see e.g. Table~1 in~\cite{Iakubovskyi:15a})
and significant distorsions of \xmm/MOS effective area in the region of our interest.
To test the sensitivity of our technique, we have simulated 5000 independent realizations of \xmm/MOS spectra 
of the Andromeda galaxy where the line was already detected~\cite{Boyarsky:14a}.
The simulations 
were performed using standard command \texttt{fakeit} inside the \texttt{Xspec} spectral fitting package.
The model parameters are set equal to best-fit model parameters of real M31 spectra seen 
by \xmm/MOS cameras~\cite{Boyarsky:14a}. The new emission line was included in a \texttt{fakeit} simulation model 
as a narrow \texttt{gaussian} model with different intensities.
For each simulation, we first modeled the obtained spectrum in \texttt{Xspec} and 
derived the new line significance $\Sigma$ using \texttt{Xspec} procedure \texttt{steppar}.
For 2 extra degrees of freedom (position and flux of the narrow line) added to our model, 
the value of $\Sigma$ and the corresponding local $p$-value (the probability of observing the extra 
line at $\sim$3.5~keV at least as extreme as that observed in simulated spectrum, 
given its absence in model spectra used for simulations, see e.g.~\cite{Babu:96,Astrostatistics-lectures}) 
can be expressed through $\Delta \chi^2$ --
the decrease of $\chi^2$ statistics when adding a narrow \texttt{gaussian} line in \texttt{Xspec} 
spectral package:
\eq{p = \frac{2}{\sqrt{\pi}}\int\limits_0^{\Sigma/\sqrt{2}}dt \exp\left(-t^2\right) = 
1-\text{e}^{-\frac{\Delta \chi^2}{2}}.\label{eq:p-value}}
After that, we processed the obtained spectrum using 
the wavelet procedure described above and derived the largest value of our wavelet paramater $S$ 
among the values of the line position $E_0$ within the energy range 3.45-3.60~keV\footnote{By doing that,
we took into account possible variations of line positions due to statistical fluctuations, see 
e.g.~\protect\cite{Iakubovskyi:15b} for details.}. To do that, we used step function wavelet with the bin 
width $W = 120$~eV and the ``mexican hat''  wavelet with $\sigma = 60$~eV. 
The obtained relation between the line significance $\Sigma$
and the value of our wavelet parameter $S$ for step function and ``mexican hat'' wavelets is plotted 
in Fig.~\ref{fig:stwsign120} and Fig.~\ref{fig:mhwsign60}, respectively.

The resulting $p$-value for $3\sigma$ line detection with our step function wavelet (calculated by 
similar simulations with no extra line added) is 0.094 corresponding to 
approximately 1.7$\sigma$ local significance. This means that wavelet method, despite its simplicity,
is able to recover 3$\sigma$ narrow lines at 1.7$\sigma$ significance.
The ``mexican hat'' wavelet shows slightly better results 
($p$-value is 0.082) but is much more time consuming, so for our quicklook analysis the step function wavelet 
is sufficient.

\section*{\sc Results and discussion}
\label{sec:results-discussion}

\indent \indent We analyzed all \xmm/MOS observations used in X-ray sky maps~\cite{Savchenko:14}
in order to search the extra narrow line at $\sim 3.5$~keV\footnote{The line position $E_0$ is allowed to 
vary within 3.45-3.60~keV range.} using step function wavelet with $W = 120$~eV described above.
The format of sky maps allowed us to combine all data from the same sky regions 
(25' $\times$ 25' squares, roughly corresponding to \xmm/MOS Field-of-View). We selected all the data
where the new line was detected at $S > 2$ (corresponding in average to $\Sigma > 4\sigma$ local significance, 
according to the best-fit line in Fig.~\ref{fig:stwsign120}).
The resulting list of 235 spatial regions is shown in Table~\ref{tab:detection-pixels}.
More detailed spectral analysis is required to reveal the presence of the line in these objects.
We leave such an analysis for future work.
 
\begin{table}
 \centering
 \caption{List of 235 $25' \times 25'$ spatial regions with the new line at $\sim 3.5$~keV detected at $S > 2$ 
 level.}
 \label{tab:detection-pixels}
 \vspace*{1ex}
  \footnotesize
 \begin{tabular}{|ccc|ccc|ccc|ccc|}
  \hline
  $S$ & RA & DEC & $S$ & RA & DEC & $S$ & RA & DEC & $S$ & RA & DEC \\
  \hline
 3.569 & 255.696 & 33.438 & 3.272 & 141.929 & -5.960 & 3.219 & 255.422 & 78.910 & 3.174 & 356.664 & -53.796 \\
 3.174 & 162.793 & 33.770 & 3.060 & 150.368 & 55.633 & 3.056 & 160.302 & 5.934 & 3.041 & 31.042 & -6.430 \\
 3.011 & 221.502 & 40.729 & 3.010 & 334.134 & -17.251 & 2.914 & 318.248 & 13.434 & 2.913 & 159.427 & 53.609 \\
 2.888 & 98.016 & -60.384 & 2.878 & 355.905 & -53.423 & 2.863 & 37.253 & 0.620 & 2.852 & 298.943 & 26.153 \\
 2.833 & 31.458 & -7.660 & 2.796 & 177.711 & -28.667 & 2.795 & 283.981 & 1.450 & 2.792 & 237.169 & 27.070 \\
 2.764 & 217.554 & 42.042 & 2.754 & 331.878 & 10.297 & 2.749 & 296.753 & 34.207 & 2.748 & 160.009 & 39.777 \\
 2.744 & 278.495 & -10.191 & 2.721 & 162.084 & -59.971 & 2.717 & 260.130 & 26.500 & 2.715 & 310.385 & -57.293 \\
 2.715 & 196.287 & -40.454 & 2.707 & 70.230 & 25.607 & 2.694 & 187.932 & 25.592 & 2.690 & 24.393 & -8.439 \\
 2.690 & 168.905 & 18.123 & 2.675 & 70.688 & 25.192 & 2.670 & 219.889 & 53.553 & 2.659 & 13.952 & -1.039 \\
 2.650 & 197.542 & 37.054 & 2.633 & 93.722 & -33.513 & 2.633 & 168.542 & 9.695 & 2.626 & 131.623 & -50.696 \\
 2.618 & 66.338 & 15.603 & 2.616 & 186.339 & 32.285 & 2.613 & 282.338 & 0.206 & 2.610 & 202.074 & -31.276 \\
 2.609 & 118.707 & 22.058 & 2.606 & 86.844 & -31.890 & 2.604 & 349.648 & -53.980 & 2.601 & 165.722 & 22.649 \\
 2.596 & 13.565 & -73.127 & 2.582 & 265.456 & -38.870 & 2.582 & 17.417 & -45.779 & 2.577 & 264.565 & 60.097 \\
 2.576 & 349.659 & -52.747 & 2.561 & 109.772 & -24.366 & 2.560 & 25.237 & -34.301 & 2.553 & 186.691 & -63.914 \\
 2.533 & 230.877 & -44.777 & 2.529 & 8.079 & 13.973 & 2.524 & 146.463 & -8.870 & 2.520 & 318.665 & 13.420 \\
 2.492 & 154.135 & -40.968 & 2.489 & 239.146 & -22.034 & 2.474 & 243.486 & -22.574 & 2.465 & 349.655 & -53.157 \\
 2.464 & 210.425 & -60.624 & 2.460 & 229.351 & -16.047 & 2.457 & 145.877 & 16.837 & 2.444 & 213.247 & 71.493 \\
 2.443 & 34.367 & -5.179 & 2.440 & 63.889 & -59.232 & 2.429 & 267.965 & 23.109 & 2.421 & 125.722 & 22.649 \\
 2.405 & 276.842 & 6.386 & 2.395 & 167.980 & 43.934 & 2.383 & 85.155 & 35.536 & 2.381 & 55.039 & -18.475 \\
 2.379 & 30.208 & -2.290 & 2.378 & 27.689 & -74.381 & 2.378 & 261.521 & 2.265 & 2.378 & 1.673 & -34.934 \\
 2.377 & 73.993 & -68.840 & 2.373 & 38.886 & -3.905 & 2.369 & 83.329 & -70.178 & 2.365 & 64.393 & 1.037 \\
 2.352 & 14.384 & -26.364 & 2.351 & 147.732 & -62.688 & 2.350 & 94.278 & 22.649 & 2.348 & 344.075 & -36.316 \\
 2.348 & 148.905 & 18.123 & 2.336 & 156.019 & -7.213 & 2.335 & 165.428 & 76.838 & 2.326 & 76.960 & -70.983 \\
 2.326 & 225.219 & 1.868 & 2.319 & 350.208 & 8.071 & 2.317 & 70.287 & -43.537 & 2.313 & 222.965 & -55.847 \\
 2.309 & 313.772 & 44.304 & 2.306 & 54.337 & -35.556 & 2.289 & 268.126 & -6.016 & 2.286 & 231.655 & 51.509 \\
 2.285 & 103.866 & -24.243 & 2.283 & 3.169 & -19.662 & 2.280 & 137.884 & 52.896 & 2.278 & 167.710 & 2.704 \\
 2.274 & 355.503 & -55.510 & 2.268 & 78.095 & -67.460 & 2.268 & 263.816 & -25.477 & 2.268 & 209.564 & -61.457 \\
 2.267 & 61.609 & -71.335 & 2.266 & 191.874 & 2.705 & 2.261 & 336.019 & -1.864 & 2.260 & 354.881 & -56.367 \\
 2.255 & 162.005 & -25.390 & 2.253 & 80.212 & -69.013 & 2.251 & 7.164 & -77.279 & 2.251 & 357.499 & 36.642 \\
 2.248 & 191.874 & 8.475 & 2.246 & 181.851 & 28.236 & 2.243 & 348.258 & -53.556 & 2.243 & 34.240 & 42.628 \\
 2.242 & 196.813 & -19.248 & 2.231 & 292.013 & 21.446 & 2.228 & 5.219 & -1.868 & 2.227 & 40.997 & -48.534 \\
 2.226 & 188.844 & 26.427 & 2.223 & 216.445 & 42.110 & 2.219 & 154.216 & -33.497 & 2.218 & 187.227 & 13.965 \\
 2.213 & 225.787 & -42.213 & 2.212 & 312.121 & 29.276 & 2.208 & 218.470 & -36.163 & 2.204 & 183.196 & 29.120 \\
 2.200 & 149.375 & 2.706 & 2.199 & 259.437 & -59.451 & 2.192 & 25.041 & -67.997 & 2.188 & 12.186 & 31.910 \\
 2.187 & 150.235 & 28.886 & 2.186 & 65.051 & 15.578 & 2.183 & 9.689 & 48.479 & 2.183 & 186.782 & -63.085 \\
 2.180 & 251.169 & 57.704 & 2.175 & 164.393 & 1.451 & 2.169 & 185.633 & 4.354 & 2.168 & 66.108 & 25.143 \\
 2.167 & 138.074 & 18.371 & 2.160 & 230.799 & -38.539 & 2.160 & 132.290 & -2.704 & 2.153 & 258.771 & -38.629 \\
 2.147 & 245.280 & -77.252 & 2.144 & 352.413 & -53.133 & 2.144 & 267.647 & -37.270 & 2.142 & 283.769 & 15.546 \\
 2.141 & 254.773 & -42.191 & 2.139 & 50.210 & 11.114 & 2.138 & 159.754 & 41.875 & 2.135 & 192.706 & 5.188 \\
 2.134 & 89.259 & -33.156 & 2.134 & 163.171 & -40.423 & 2.133 & 139.292 & 46.464 & 2.132 & 333.952 & 0.208 \\
 2.132 & 308.751 & -33.974 & 2.131 & 68.988 & -78.124 & 2.131 & 244.312 & 12.279 & 2.128 & 128.854 & 25.190 \\
 2.123 & 265.687 & -23.892 & 2.121 & 263.285 & -33.798 & 2.116 & 263.813 & -33.415 & 2.115 & 20.302 & -0.205 \\
 2.112 & 26.911 & 61.840 & 2.111 & 227.306 & 57.265 & 2.111 & 157.884 & 30.873 & 2.109 & 341.392 & 28.208 \\
 2.108 & 67.006 & 25.989 & 2.107 & 18.820 & -47.323 & 2.102 & 181.945 & -32.489 & 2.100 & 23.563 & -36.290 \\
 2.099 & 291.067 & 13.978 & 2.098 & 136.842 & 0.621 & 2.097 & 8.103 & 39.776 & 2.095 & 39.196 & 61.565 \\
 2.093 & 179.233 & 52.798 & 2.086 & 86.087 & -25.967 & 2.084 & 50.652 & 16.877 & 2.082 & 255.367 & 59.682 \\
 2.082 & 177.689 & -28.263 & 2.082 & 13.019 & 27.220 & 2.080 & 147.337 & 76.449 & 2.079 & 314.896 & 43.847 \\
 2.079 & 165.122 & -77.666 & 2.079 & 157.829 & -34.967 & 2.079 & 140.755 & 30.379 & 2.079 & 122.056 & -76.337 \\
 2.073 & 64.594 & 29.184 & 2.073 & 37.289 & -29.498 & 2.070 & 351.883 & -10.702 & 2.069 & 248.988 & 78.124 \\
 2.065 & 77.516 & -69.129 & 2.063 & 220.125 & 64.432 & 2.061 & 14.866 & -72.689 & 2.060 & 348.913 & 78.957 \\
 2.059 & 40.445 & -59.862 & 2.057 & 127.269 & -33.539 & 2.056 & 272.428 & -19.359 & 2.055 & 195.897 & -83.920 \\
 2.053 & 200.159 & -63.615 & 2.050 & 178.886 & 6.761 & 2.049 & 10.208 & -9.293 & 2.047 & 58.479 & -0.206 \\
 2.047 & 179.419 & 26.530 & 2.043 & 34.367 & -6.823 & 2.042 & 93.492 & -27.619 & 2.042 & 237.051 & -32.141 \\
 2.042 & 144.735 & 41.338 & 2.041 & 154.435 & -58.882 & 2.039 & 89.482 & -66.430 & 2.035 & 304.006 & 37.142 \\
 2.033 & 230.222 & 20.208 & 2.031 & 52.140 & 30.285 & 2.029 & 281.929 & -3.091 & 2.027 & 133.227 & 33.527 \\
 2.027 & 103.129 & 40.837 & 2.025 & 182.005 & 25.390 & 2.021 & 163.093 & 35.850 & 2.018 & 321.130 & 51.185 \\
 2.016 & 325.166 & -43.020 & 2.014 & 323.903 & -54.653 & 2.014 & 243.469 & -22.987 & 2.014 & 181.895 & -27.427 \\
 2.009 & 86.957 & -70.276 & 2.006 & 35.194 & -4.761 & 2.004 & 267.710 & -6.837 & 2.002 & 80.552 & -68.144 \\
 2.002 & 14.723 & -66.772 & 2.001 & 67.169 & -17.274 & 2.000 & 83.981 & -4.754 &       &        &         \\
 \hline 
 \end{tabular}
\end{table}

\section*{\sc acknowledgement}
\indent \indent We thank the referee, Dr.~Oleksiy Agapitov, 
for the comments that significantly improved the quality of the paper.
This work has been partially supported from the Swiss National Science Foundation 
grant SCOPE IZ7370-152581, the grant No.~F64/42-2015 of the State Fund for Fundamental Research of Ukraine, 
the Program of Cosmic Research of the National Academy of Sciences of Ukraine, 
and the State Programme of Implementation of Grid Technology in Ukraine.


\let\jnlstyle=\rm\def\jref#1{{\jnlstyle#1}}\def\aj{\jref{AJ}}
  \def\araa{\jref{ARA\&A}} \def\apj{\jref{ApJ}\ } \def\apjl{\jref{ApJ}\ }
  \def\apjs{\jref{ApJS}} \def\ao{\jref{Appl.~Opt.}} \def\apss{\jref{Ap\&SS}}
  \def\aap{\jref{A\&A}} \def\aapr{\jref{A\&A~Rev.}} \def\aaps{\jref{A\&AS}}
  \def\azh{\jref{AZh}} \def\baas{\jref{BAAS}} \def\jrasc{\jref{JRASC}}
  \def\memras{\jref{MmRAS}} \def\mnras{\jref{MNRAS}\ }
  \def\pra{\jref{Phys.~Rev.~A}\ } \def\prb{\jref{Phys.~Rev.~B}\ }
  \def\prc{\jref{Phys.~Rev.~C}\ } \def\prd{\jref{Phys.~Rev.~D}\ }
  \def\pre{\jref{Phys.~Rev.~E}} \def\prl{\jref{Phys.~Rev.~Lett.}}
  \def\pasp{\jref{PASP}} \def\pasj{\jref{PASJ}} \def\qjras{\jref{QJRAS}}
  \def\skytel{\jref{S\&T}} \def\solphys{\jref{Sol.~Phys.}}
  \def\sovast{\jref{Soviet~Ast.}} \def\ssr{\jref{Space~Sci.~Rev.}}
  \def\zap{\jref{ZAp}} \def\nat{\jref{Nature}\ } \def\iaucirc{\jref{IAU~Circ.}}
  \def\aplett{\jref{Astrophys.~Lett.}}
  \def\apspr{\jref{Astrophys.~Space~Phys.~Res.}}
  \def\bain{\jref{Bull.~Astron.~Inst.~Netherlands}}
  \def\fcp{\jref{Fund.~Cosmic~Phys.}} \def\gca{\jref{Geochim.~Cosmochim.~Acta}}
  \def\grl{\jref{Geophys.~Res.~Lett.}} \def\jcp{\jref{J.~Chem.~Phys.}}
  \def\jgr{\jref{J.~Geophys.~Res.}}
  \def\jqsrt{\jref{J.~Quant.~Spec.~Radiat.~Transf.}}
  \def\memsai{\jref{Mem.~Soc.~Astron.~Italiana}}
  \def\nphysa{\jref{Nucl.~Phys.~A}} \def\physrep{\jref{Phys.~Rep.}}
  \def\physscr{\jref{Phys.~Scr}} \def\planss{\jref{Planet.~Space~Sci.}}
  \def\procspie{\jref{Proc.~SPIE}} \let\astap=\aap \let\apjlett=\apjl
  \let\apjsupp=\apjs \let\applopt=\ao \def\jcap{\jref{JCAP}}
\providecommand{\href}[2]{#2}\begingroup\raggedright\endgroup

\end{document}